\documentclass[epj,referee]{svjour}
\usepackage{graphics}
\usepackage{bm,bbm,amsmath,amssymb,color,subeqnarray,graphicx,dcolumn}
\usepackage{graphicx,epsfig,latexsym,float,subfloat,subfigure,ulem,gensymb,varwidth,booktabs}
\usepackage{comment}
\usepackage{cite}
\usepackage{amsmath}
\usepackage{balance}
\usepackage{color}
\usepackage{soul}
\usepackage{hyperref}
\begin{document}
\title{Optimal observables to determine  entanglement of a two qubit
 state}

\author{Gaurav Chaudhary\inst{1,2}\thanks{\emph{e-mail:}gaurav@physics.utexas.edu } \and V. Ravishankar\inst{2,3}
%
}                     
\institute{Department of Physics, The University of Texas at Austin, TX-78712 (USA) \and Department of Physics, Indian Institute of Technology, New Delhi-110016 (India) \and Department of Physics, Indian Institute of Technology, Kanpur-208016(India)}

\date{Received: date / Revised version: date}
%
\abstract{
Experimental determination of entanglement is important not only to characterize the  state and use it in quantum information, but also in understanding complicated phenomena such as phase transitions. In this paper we show  that in many cases, it is possible to determine entanglement of a two qubit state, as represented by concurrence, with a few observables, {\it most of which are local}. In particular, rank 1 and rank 2 states need exclusively  measurement of local observables while rank 3 states need measurement of just one correlation observable in addition to local observables. Only the rank 4 states are shown to require a more detailed tomography. The analysis also sheds light on the other measure, non separability since it is a lower bound on concurrence.
\PACS{
      {PACS-key}{discribing text of that key}   \and
      {PACS-key}{discribing text of that key}
     } 
} 
\maketitle

\section{Introduction}
\label{intro}
The purpose of this paper is to examine the possibility of determining  entanglement of two qubit states without exhausting the entire observable space. We look for  minimal sets  of observables that can specify the entanglement of a given  state.  An answer to this question would not only bring out important fundamental aspects of entanglement characterization but also in specific cases can save an experimentalist of measurement of unwanted extra observables that do not effect entanglement independently-- which is by no means a trivial task. It will also bring out the invariant character of entanglement with respect to a host of transformations, especially local transformations. Since the interest in entanglement has now transcended the domain of quantum information and is being used in understanding varied physical phenomena, ranging from black hole thermodynamics to phase transitions,  it is all the more   pertinent to identify sets of observables which can decisively yield information on entanglement. Finally, not all observables can be measured with equal ease. For instance, measuring local observables is much easier than measuring correlations, and it would be good to employ easily measurable observables.

This topic is not new and has been addressed earlier in a variety of contexts. Experimentally, tomography has been the natural route to determine entanglement. Tomographic techniques have been well developed and applied for different qubit systems, viz. superconducting qubits \color{blue}\cite{Filipp, Steffen}\color{black}, entangled photons \color{blue}\cite{Vasilyev}\color{black}. The techniques employed depend on the system considered, and are well summarized recently  in \color{blue}\cite{bana}\color{black}. Tomography involves multiple measurements on identical copies and constructing density operator \color{blue}\cite{agnew, Chiuri, LinPeng, Babichev, Roos}\color{black}. We may also mention, as examples, tomography of photons entangled in high dimensions \color{blue}\cite{agnew}\color{black}. A complete state tomography requires measurement of a complete set of observables, which is essentially equal to number of elements in density operator, hence the complexity scales exponentially with dimensionality of quantum system \color{blue}\cite{agnew, Roos}\color{black}. Indeed, some cases of tomography have been reported, which require fewer observables, but these cases are either limited to approximation schemes \color{blue}\cite{Gross}\color{black}, or rather very specific quantum states \color{blue}\cite{Rehacek}\color{black}. We refer to \color{blue}\cite{Cinelli, Souza, Altepeter, Toth, Fei, Walborn, Walborn2} \color{black} which employ such specific states.  Theoretical investigations have been conducted to reconstruct density matrix efficiently using numerical methods \color{blue}\cite{James}\color{black}. 

The strategy employed in this paper is to analyze the states rankwise.  Similar approaches have been employed successfully earlier \color{blue}\cite{Horodecki, Guhne, Guhne2, Iman,Yong, Li, KB1, KB2}\color{black}; however, there are  differences between those works and ours  in several aspects. Some of them employ  numerical or nondeterministic approach \color{blue}\cite{Iman, Yong}\color{black}. The very interesting result in \color{blue}\cite{Horodecki} \color{black} that measurements on very few observables are required to determine entanglement of formation, requires joint measurements of subsystems while our approach focusses on local measurements unless absolutely is necessary. Recent works by Bartkiewicz et al. \color{blue}\cite{KB1, KB2} \color{black} are closely related to our approach of using local invariants to quantify entanglement. The approaches taken by them and by us complement each other since we look at two related but independent measures of entanglement. While we focus on concurrence as an entanglement measure, the focus there is on negativity, which is lower bound on concurrence. Universal entanglement witness (UWE) discussed in \color{blue}\cite{KB2} \color{black} is defined as the determinant of a partially transposed density matrix. Our approach unlike UWE (which provides tight bound on entanglement), leads to full determination of entanglement in most cases. Our demonstrations are constructive and we employ the full machinery of invariance under local transformations. Earlier,  Yang et al. \color{blue}\cite{Yang} \color{black} have discussed two qubit entanglement in terms of invariants of a parent three qubit systems. We restrict our discussion to the  invariants of 2QS, getting rid-off any unnecessary measurements involving more qubits than required. Hence the invariants in our discussion require measurements only on the qubits of which entanglement is being discussed.

\section{Entanglement measures}
\label{sec:1}
We first need to specify a measure of entanglement before identifying the relevant observables. We consider the pure states first, partly for completeness and partly to provide the setting.

\subsection{Pure states}
Let $\vert \Psi \rangle$ be a state of a bipartite system with subsystems $A,B$. Its entanglement, $E(\Psi )$ could be specified in many ways such as violation of Bell's inequalities,  or the entropy of one of the reduced states. Fortunately, all such characterizations are equivalent in the sense that they are relative monotones of each other. In this paper,  we  choose pure state concurrence as our measure since it easily generalizes to the 2QS. It can  be conveniently expressed as $C(\Psi) = 2\vert \langle \tilde{\Psi}|\Psi \rangle \vert$ where  $|\tilde{\Psi}\rangle$ is its time reversed state. More explicitly, if we employ the language of spin -- which we do throughout for convenience, writing
 $\vert \Psi^{AB} \rangle = \sum_{m\mu}c_{m\mu}\vert m,\mu \rangle; ~~ m, \mu =\pm 1/2$, 
the time reversed state is given by
$
\vert \tilde{\Psi} \rangle  = \sum_{m \mu} (-1)^{m +\mu}c^{*}_{-m,-\mu}\vert m,\mu\rangle.
$
Thus the expression for pure state concurrence is given by
\begin{equation}
C(\Psi) = 2 \vert c_{1/2,1/2}c_{-1/2,-1/2} - c_{1/2,-1/2}c_{-1/2,1/2}\vert
\label{pc}
\end{equation}
The factor 2 in the above expression bounds the measure in the interval $[0,1]$.

Before we proceed further, it is pertinent to note that any measure of entanglement is required to be invariant under local transformations, $SU(2) \times SU(2)$ which is a proper subgroup of $SU(4)$. Since there are only two independent local invariants for a pure state, its norm being one, it follows that all measures are guaranteed to be equivalent.

\subsection{Mixed state measures}
The case with mixed states is more involved for several reasons.
(i) It is first necessary to fix the measure of entanglement that we wish to employ for mixed states. For, unlike the pure case, there is no unique measure of entanglement. Several measures that have been proposed \color{blue}\cite{werner, wootters, nielsen, bhardwaj} \color{black} are all inequivalent to each other; they are not relative monotones of each other.  Indeed, a mixed state can have as many as nine local invariants, and any measure of entanglement is a complicated function of these invariants. (ii) Even if one imagines that nine measurements should, therefore, suffice, it may not be the case because experimental techniques are generally geared to measure single particle observables and correlations. Even theoretically, expressing a measure of entanglement in terms of a standard set of invariants is a tedious task. Consequently, it is worthwhile looking for sets of optimal measurements which can determine entanglement maximally.

The choice of observables naturally depends on the measure of entanglement. In this paper we employ
concurrence which naturally generalizes pure state concurrence \color{blue}\cite{wootters}\color{black}. It has a rather involved 
algebraic expression given by \color{blue}\cite{hill}\color{black}

\begin{equation}
C(\rho)=max(0,{\lambda_1}-{\lambda_2}-{\lambda_3}-{\lambda_4})
\label{conc}
\end{equation}
where $\lambda_1$, $\lambda_2$, $\lambda_3$, $\lambda_4$ are the square roots of eigenvalues of ${\rho}{\tilde{\rho}}$ in decreasing order and $\tilde{\rho}$,   the spin flipped  density operator is essentially the time reversed state of $\rho$.

\section{Local invariants}
\label{sec:3}
Let the state be written in its standard form
\begin{equation}
\rho^{AB}_{\Psi} = \frac{1}{4} \Big\{ 1 + \vec{\sigma}^A\cdot \vec{P} + \vec{\sigma}^B\cdot \vec{S} 
 + \sigma^A_i\sigma^B_j \Pi_{ij} \Big\},
\label{state}
\end{equation}
It is convenient to further define, in addition,  
\begin{eqnarray}
 A_i = \Pi_{ij}P_j;&  ~ &  B_i = S_j\Pi_{ji} \nonumber \\
\vec{\alpha} = \vec{P} \times \vec{A} &;&  \vec{\beta} = \vec{S} \times \vec{B} \nonumber \\
T_{ij} =  \Pi_{ik}\Pi_{jk} & & 
\end{eqnarray}
A single-qubit operation is represented by two orthogonal transformations $R_{1}, R_{2} \in SU(2)$ on respective quits $A$ and $B$, which transforms $\rho^{AB}_{\Psi} $ as rule:  $\vec{P} \to R_{1}\vec{P}$, $\vec{S} \to R_{2}\vec{S}$ and $\Pi_{ij} \to R_{1(ik)} \Pi_{kl} R^{T}_{2(jl)}$. Based on this a set of independent $SU(2) \times SU(2)$ invariants can then be conveniently listed as follows:
\begin{eqnarray}
I_1 = \vec{P}^2&; &  I_2 = \vec{S}^2 \nonumber \\
I_4 = \vec{A}^2 &;&  I_3= \vec{P}\cdot \vec{A}=\vec{S}\cdot \vec{B}
 \nonumber \\ 
I_5 = \vec{B}^2 & ;& I_6 = Tr (T) \nonumber \\
 I_8 = Tr (T^2) & ; & I_7 = A_i\Pi_{ij}B_j\nonumber \\
I_9 &= &\alpha_i\Pi_{ij}\beta_j.
\label{inv}
\end{eqnarray}

Since entanglement is invariant under local transformations, it would appear that even in the most general case, a measurement of the above nine invariants should suffice. In fact, concurrence is invariant for states under much larger set of transformation. For example, concurrence vanishes for  a pure separable state (of rank 1) , any three dimensional projection (which is of rank 3)  and of course the fully unpolarized state (of rank 4)- one may even expect a further decrease in the number of measurements.  However, such a counting is rather too simplistic; one only need to look at the form of these invariants in Eq. \color{blue}\ref{inv}\color{black}. It would be  a tall order to ask that  experiments be  designed to measure these invariants directly. 

With this in mind we approach the problem by examining specific classes of states with increasing degree of complexity. We choose the rank of the state, $R(\rho)$ to characterize the state. Finer distinctions will be made for each rank as and when possible.

\section{Entanglement of  pure states ($R(\rho)=1$)}
\label{sec:4}
Well known  though this case is, a brief discussion would highlight several features which can be exploited for  the more complicated mixed states.

\subsection{Determination of $C(\Psi)$}
The expression for pure state concurrence is given in Eq. \color{blue}\ref{pc}\color{black}. To determine the observables, it is
better to rewrite it in the form given in Eq. \color{blue}\ref{state}\color{black}.
 The requirement of purity, $\rho^2 =\rho$, imposes additional conditions
\begin{eqnarray}
\vec{P}^2  & =  & \vec{S}^2 \nonumber \\
2 \vec{P}^2  + Tr (T)  & = & 3.
\end{eqnarray}
Well known that these relations are, it is still noteworthy that the strength of the correlation, given by
$Tr (T)$,  is entirely determined by the single qubit invariant, $\vec{P}^2$. Thus, although entanglement depends on correlation, it can more easily  be determined by measuring a simpler observable, {\it viz.}, of the degree of polarization, $\vec{P}^2$ of either of the subsystems. Indeed, concurrence is simply given by $C(\Psi) = \sqrt{1 - \vec{P}^2}$. In short,  a local observable determines the nonlocal character of the state unambiguously. One hopes that similar opportunities present themselves for mixed states as well.

Walborn et al. \color{blue}\cite{Walborn, Walborn2} \color{black}have already provided more efficient methods to determine entanglement, which apply only for pure state, while we systematically extend our studies to mixed states. We discuss each case on the basis of rank of the density operator and also discuss special examples.

\section{Rank two states}
\label{sec:5}

Rank two states occur naturally  as descendants over a three qubit pure state. Physical realizations  abound.  The final states in the celebrated beta decay, $n \rightarrow p + e^- +\bar{\nu}$ is one such example. 
Monogamy relations \color{blue}\cite{kundu} \color{black}make the concurrences of the three 2-qubit subsystems mutually constraining. 

 Let us begin with  a convenient representation of  a rank 2 state in its eigenbasis:

\begin{equation}
{\rho_2}={\nu}|\chi\rangle\langle\chi|+(1-{\nu})|\chi_{\perp}\rangle\langle\chi_{\perp}\vert.
\label{r2}
\end{equation}
It is further convenient to employ the Schmidt basis to expand $\vert \chi \rangle$, and employ the residual freedom in local operations, together with the overall phase,  to simplify the form of $\vert \chi_{\perp} \rangle$. Thus,

\begin{eqnarray}
|\chi\rangle & =&  \cos\alpha |00\rangle+\sin \alpha|11\rangle \nonumber \\
  |\chi_{\perp}\rangle & = & \cos\eta\big\{\sin \alpha |00\rangle -\cos\alpha|11\rangle\big\} \nonumber \\
&+& \sin \eta \big\{ \sin \beta \exp(-i\gamma) |01\rangle 
 +\cos \beta |10\rangle \big\}
\label{r2l}
\end{eqnarray}
where $0 \le \alpha, \beta, \eta  \le \pi/2$ and $0 \le \gamma \le 2\pi$. Thus the manifold of inequivalent rank 2 states (under LO) is characterized by a family of five parameters, $\nu, \alpha, \beta, \gamma, \eta$. Furthermore, concurrence is a function of only local invariants.  It remains to relate them to easily measurable observables.

\noindent {\it Claim I}: Concurrence of a rank 2 state is completely determined by local measurements except for some values of parameters for which the equations become degenerate\footnote{These degeneracy points are clearly a set of measure zero of an observable in Eqs. \color{blue}\ref{pa},\ref{pb}\color{black}. We discuss one such example in Sec. \color{blue}\ref{sec:5.2}\color{black}}. 

To show this, we prove a stronger result: \\
\noindent {\it Claim $I^{\prime}$}:  A two qubit rank 2 state is completely determined by its daughter one qubit states, upto local transformations. 

The proof is by explicit construction.  The local observables  can be read off from Eq. \color{blue}\ref{r2l} \color{black}to be
\begin{eqnarray}
P_x  &= & 2(1-\nu)\cos\eta\sin\eta\big\{ \cos\alpha \cos\beta + \sin \alpha \sin\beta\cos\gamma\big\} \nonumber \\
P_y&= & -2(1-\nu)\cos\eta\sin\eta\sin\alpha \sin\beta\sin\gamma\nonumber \\
P_z &= & \nu\cos2\alpha+(1-\nu)\big\{-\cos^2\eta \cos2\alpha \nonumber \\ &&+\sin^2\eta\cos2\beta\big\}
\label{pa}
\end{eqnarray}

Similarly the expression for $\vec{S}$ can be read off as 
\begin{eqnarray}
S_x &= & 2(1-\nu)\cos\eta\sin\eta\big\{-\sin\alpha \cos\beta+\sin\alpha\sin\beta\cos\gamma \big\}  \nonumber \\
S_y &= & -2(1-\nu)\cos\eta\sin\eta\cos\alpha\sin\beta\sin\gamma \nonumber  \\
S_z &= & \nu\cos2\alpha-(1-\nu)\{\cos^2\eta\cos2\alpha  \nonumber  \\
&&+ \sin^2\eta\sin2\beta\}
\label{pb}
\end{eqnarray} 

Note that $\sin\alpha$ and $\cos\alpha$ determine each other as is the case with $\sin\beta$, $\cos\beta$ and also $\cos\eta$, $\sin\eta$. . However, one needs to measure both  $\cos\gamma ~ \rm{and} ~ \sin\gamma$ for determining $\gamma$. Thus measurement of six observables are required  to determine the state, though the manifold is itself five dimensional.

It is easy to see from Eqs. \color{blue}\ref{pa},\ref{pb} \color{black}that the two single qubit polarizations determine the state completely. We outline the sequential steps below:
\begin{enumerate}
 \item  The ratio $P_y/S_y$ determines $\alpha$ unambiguously. 
\item  We next note that the ratios $P_x/S_x  \equiv R_1(\beta,\gamma)$ and $P_y/P_x  \equiv R_2(\beta,\gamma)$ together determine $\beta,\gamma$. There would still be  a discrete ambiguity in $\gamma$.
\item Combining them with $P_z \pm S_z$, $\nu, \beta$ are also determined. 
\item Finally, the discrete ambiguity in the value of $\gamma$ can be resolved by employing either $P_y$ or $S_y$.
\end{enumerate}
We have thereby proved that rank 2 density matrix,  upto its local equivalents, can be constructed by single qubit observables except for some cases Sec. \color{blue}\ref{sec:5.2}\color{black}.

Concurrence is a function of local invariants, and they are simply given by $I_1, I_2$ in this case. We have thus shown that  a measurement of two single qubit invariants is necessary, and sufficient to determine $C(\rho)$ whenever it is of rank 2. In the limiting case when $\nu =0,1$, the two invariants are equal, thereby reducing to the pure case discussed in the previous section.

To bring out vividly the result proved, we consider several examples which are based on another equivalent representation of rank 2 states, as an incoherent superposition of a separable and a pure state:
\begin{equation}
\rho_2 = \lambda \rho_s + (1-\lambda)\vert \psi \rangle \langle \psi \vert
\label{sep}
\end{equation}
where the rank of the separable state, $R(\rho_s) \le 2$.

Accordingly, we consider two cases, $R(\rho_s) =1,2$ separately. We first discuss the latter case, and then go on to show the exceptional nature of rank 1 states.

\subsection{$R(\rho_s)=2$}
The most general separable state, $\rho_s$ is a mixture of two separable and mutually orthogonal states: 

\begin{equation}
\rho_{s}=\mu|\chi_1\rangle\langle\chi_1|+(1-\mu)|\chi_2\rangle\langle\chi_2|
\label{sep2}
\end{equation}
Employing local transformations, we can always write $|\chi_1{\rangle} \\ =|00\rangle$ and $|\chi_2{\rangle}=a|10\rangle+b|11\rangle$ where $a,b$ can be taken to be real and nonnegative. Note that $\vert \psi \rangle$ necessarily lies in the plane spanned by $\vert \chi_1 \rangle$ and $\vert \chi_2 \rangle$, and may be written as

\begin{equation}
|\psi\rangle=\cos\theta|00\rangle+\exp({i\beta})\sin{\theta}(a|10\rangle+b|11\rangle)
\label{ms}
\end{equation}

Combining Eqs. \color{blue}\ref{sep}\color{black}, \color{blue}\ref{sep2} \color{black}and \color{blue}\ref{ms} \color{black}we get the canonical form of the state. The basic algebraic steps to determine concurrence are:
\begin{enumerate}
\item Construct the spin flip density operator $\tilde{\rho_2}$ and compute $\rho_2\tilde{\rho_2}$.
\item Solving for eigenvalues of $\rho_2\tilde{\rho_2}$ becomes simpler due to the fact that two of the eigenvalues vanish.
\item Once eigenvalues are known it's straightforward to write concurrence as a function of elements of the density operator and further as a function of local invariants.\footnote{The algebra is long but straightforward and in parts involves some clever substitutions. The basic procedure for all this work is same; once concurrence is written in terms of elements of density operator, we write expectation values of observables and from observables we construct the invariants and write them as a function of elements of density operator. Now we substitute invariants for the density operator elements in concurrence expression.} 
\end{enumerate}
The final expression is quite neat, and is given by

\begin{equation}
C=max(\sqrt{1-I_1}, \sqrt{1-I_2})
\label{Conc.R2}
\end{equation}
The above expression is similar to the pure state case, but while pure state entanglement can be determined by local measurements on only one of the qubit, entanglement of this case requires local measurements on both qubits.

\subsection{$R(\rho_{s})=1$ }
\label{sec:5.2}   
We use this case to give an example of a rank 2 state that lies in the degeneracy points mentioned in {\it Claim I}, which are exception to the rule proven in Sec. \color{blue}\ref{sec:5}\color{black}. For that, we consider a special case when the two one dimensional projections are orthogonal to each other.
Again, using the freedom under local transformations, we may employ the forms
\begin{eqnarray}
 \rho_s & = & \vert 00\rangle\langle 00\vert \nonumber \\ 
\vert \psi \rangle  & =  & r_1\vert 01\rangle + c\vert 10\rangle + r_2\vert 11\rangle \nonumber \\
\rho_2 & \equiv & \lambda \vert 00 \rangle \langle 00 \vert + (1-\lambda)\vert \psi\rangle\langle \psi \vert
\label{1s}
\end{eqnarray}
where $r_{1,2}$ are real and nonnegative.

The evaluation of concurrence is again straight forward and all the steps involved are similar to the last case.
It has a very simple form given by

\begin{equation}
C=(1-\lambda)2r_1|c| \equiv (1-\lambda)C(\vert \psi\rangle)
\label{r1}
\end{equation}.

Simple though the expression is, it is clear that single qubit invariants are not sufficient since the determination of the eigenvalue $\lambda$ requires a knowledge of tensor correlations as well. Such states necessitate  a determination of an additional quantity such as $Tr(\rho^2)$ which is a measure of mixedness of $\rho_2$: $Tr({\rho_2}^2)=2{\lambda}^2-2\lambda+1$. Here we have proved that for this class of quantum states entanglement can be specified in terms of single qubit invariants and mixedness.

\subsubsection{Two dimensional projections}
To illustrate the above remark physically, we consider the special case when the state is a two dimensional projection. This corresponds to $\lambda=\frac{1}{2}$ in Eq. \color{blue}\ref{1s}\color{black}. Since the numerical value of mixedness has been fixed, the concurrence should only depend on single qubit invariants. It is easy to verify that  concurrence can be expressed in terms of the single qubit invariants as 

\begin{equation}
C=\sqrt{\frac{1-I_1-I_2}{2}-\sqrt{\frac{(I_1-I_2)^2}{4}-\frac{I_1+I_2}{2}+\frac{1}{4}}}
\label{intinv}
\end{equation}

\subsubsection{A physical example}
We show how the preceding analysis is applicable to the entanglement studied for a physical system involving quantum phase transition \color{blue}\cite{bose, Wu}\color{black}, where QPT is evident in non-analyticity of entanglement. The system being studied is frustrated two-leg spin-1/2 ladder, with the Hamiltonian given by $H_{ladder}=\sum\nolimits_{\langle ij \rangle}J_{ij}\vec{S_{i}}.\vec{S_{j}}-h\sum\limits_{i=1}^n S^z_{i}$. The exchange interaction along the rungs is $J_{ij}=J_R$ and both the intra-chain nearest-neighbor and diagonal exchange interactions are $J_{ij}=J$. This model shows 1QPTs at $h_{c1}$ and $h_{c2}$. In the limit $N\rightarrow\infty$, the ground state for $h<h_{c1}$, is a tensor product of singlet along the rungs and for $h>h_{c2}$ ground state is a tensor product of all rungs in unentangled triplet state. These states are example of pure states with maximal entanglement and separable states respectively. The ground state of the system in $h_{c1}<h<h_{c2}$ is a statistical mixture of singlet and triplet states with equal probability. It is straightforward to identify this to be an example of two dimensional projection in this regime of $h$. It follows from our analysis that measuring concurrence for this system is relatively easy! Density operator naturally  has the form given in Eq. \color{blue}\ref{1s} \color{black}with $r_2=0$, $r_1=|c|=\frac{1}{\sqrt{2}}$.  It is easy to see that concurrence for this state is simply given by
\begin{equation}
C=2(1-\lambda)r_1|c| \equiv 1-\rho_{11}
\end{equation}
in a convenient measurement basis. Similarly, the invariant character of concurrence is manifested by Eq. \color{blue}\ref{intinv}\color{black}. This is one of the few examples where a single occupation probability $\rho_{11}$ determines entanglement completely. Value of $\rho_{11}$ can be picked out by the projective measurement of correlation observable as $\frac{1}{2}(\langle s_zp_z\rangle+1)$.

\section{Rank three states} 
\label{sec:6}
Rank 3 states are considerably more complex, and as we show below, measurement of correlations is inevitable in most situations. Yet  we may still ask how far one can take the program of minimal measurements through. To examine that, let us write the state in its eigen basis:

\begin{equation}
\rho_3=\nu_1|\chi_1\rangle\langle\chi_1|+\nu_2|\chi_2\rangle\langle\chi_2|+(1-\nu_1-\nu_2)|\chi_3\rangle\langle\chi_3|
\end{equation}
where we have arranged the eigenvalues $\nu_i$  in the nondecreasing order. The eigenstates $\vert\chi_i\rangle$ may be brought to their canonical form under local transformations. We may choose the first two eigenstates to be a given in Eq. \color{blue}\ref{r2l} \color{black}and write the third eigenstate as 

\begin{eqnarray}
|\chi_1\rangle & =&  \cos\alpha |00\rangle+\sin \alpha|11\rangle \nonumber \\
  |\chi_2\rangle & = & \cos\eta\big\{\sin \alpha |00\rangle -\cos\alpha|11\rangle\big\} \nonumber \\
&+& \sin \eta \big\{ \sin \beta \exp(-i\gamma) |01\rangle 
 +\cos \beta |10\rangle \big\} \nonumber \\
 |\chi_3\rangle & = & \cos\xi\big\{\sin \alpha |00\rangle -\cos\alpha|11\rangle\big\} \nonumber \\
&+& \sin \xi \big\{ \sin \theta \exp(-i\phi_1) |01\rangle
+ \cos \theta \exp(-i\phi_2)|10\rangle \big\} \nonumber \\
\label{rho3}
\end{eqnarray}
The orthogonality conditions impose two further constraints on the angles. Thus the state, in its canonical form, has eight independent parameters. 

An immediate corollary is that local measurements are not sufficient to determine concurrence and must be necessarily supplemented with  measurements of correlations. The same feature is inherited by rank 4 states, since they would be characterized by the maximal number of parameters, viz, nine, in their canonical form\footnote{Six out of the fifteen parameters can be transformed away by local transformations}.While there could be special cases where local measurements may suffice, we consider, instead an alternative problem: that of extraction of maximal information on concurrence with fewer observables. That is necessarily of the nature of placing bounds, and is closer in spirit to the approach of Plenio \color{blue}\cite{wunderich, audenaert}\color{black}. But before that we look at a class of physical systems which is described by a rank 3 density operator.

\subsection{A physical example} 
The density operator we study occurs physically in spin pair in AKLT state \color{blue}\cite{bose, aklt}\color{black},  two neighboring sites in antiferromagnetic XXZ chain \color{blue}\cite{Gu} \color{black}and Dicke state \color{blue}\cite{Wang} \color{black}and is of the form
\begin{equation}
{\rho}=
\begin{bmatrix}
u_{+} & 0 & 0 & 0\\
0 & w_{1} & z & 0\\
0 & z^{*} & w_{2} & 0\\
0 & 0 & 0 & u_{-}              
\end{bmatrix}
\label{r3p}
\end{equation}
The entanglement of such states is well studied and is given by $C=2\, max(0, |z|-\sqrt{u_{+}u_{-}})$\color{blue}\cite{O'connor}\color{black}. Although entanglement measurements of such systems requires measurement of correlation observables, it is important to bring out the invariant character of the entanglement. The algebraic approaches applied in previous examples are followed to get the expression for concurrence 
\begin{equation}
C=\sqrt{2(I_8-\frac{I_5}{I_1})}-\sqrt{(1+\sqrt{\frac{I_5}{I_1}})^2-(I_1+I_2)^2}
\label{c3i}
\end{equation}

\subsection{Bounds on concurrence}
As mentioned earlier we study bounds on concurrence. With a slight rearrangement we can write a rank 3 state as $\rho_3=\lambda\Pi_3+(1-\lambda)\rho_2$  where $\Pi_3$ is a three dimensional projection and $R(\rho_2) =2$. Further employing the resolution 
in Eq.\color{blue} \ref{sep} \color{black} for $\rho_2$, we get
\begin{equation}
\rho_3=\lambda\Pi_3+(1-\lambda)[\mu\rho_{sep}+(1-\mu)|\psi\rangle\langle\psi|]
\label{r3}
\end{equation}

The only contribution to entanglement of the state described by $\rho_3$ in Eq. \color{blue}\ref{r3} \color{black}is from the pure state component,  $|\psi\rangle$ since the other terms are concurrence free. Thus, we have a rather weak bound
\begin{equation}
C(\rho_3)\leq(1-\lambda)(1-\mu)C(\psi)\leq(1-\lambda)C(\psi)
\label{ineq.1}
\end{equation}
The pure state, $|\psi\rangle$ is constrained to to  lie in the three dimensional  space $\mathcal{H}_3$ projected by $\Pi_3$.
\begin{align}
\begin{split}
&{|\psi\rangle}=\cos\theta(a|01\rangle+b|10\rangle) \\
&+\sin{\theta}\cos\phi|00\rangle+\sin{\theta}\sin\phi|11\rangle
\end{split}
\end{align}
An interesting question is to find the upper limit on $\lambda$ for the projection $\Pi_3$ for a state to be entangled. That  happens when the weight associated with $\rho_{sep}$ vanishes, and the pure component is fully entangled. The phases are not of much concern for this maximally entangled state, and thus, 
\begin{equation}
{\rho^{max}_3}=
\begin{bmatrix}
\frac{\lambda}{3} & 0 & 0 & 0\\
0 & (1-\frac{2\lambda}{3})a^2 & (1-\frac{2\lambda}{3})ab & 0\\
0 & (1-\frac{2\lambda}{3})ab & (1-\frac{2\lambda}{3})b^2 & 0\\
0 & 0 & 0 & \frac{\lambda}{3}              
\end{bmatrix}
\label{maxent1}
\end{equation}
for which, the concurrence is given by
\begin{equation}
C(\rho^{max}_3)=2(1-\frac{2\lambda}{3})ab-\frac{2\lambda}{3}
\end{equation}
which implies that for entanglement
\begin{equation}
\lambda<\frac{3ab}{1+2ab}
\label{bound1}
\end{equation}
The R.H.S. of above expression attains its maxima at $ab=\frac{1}{2}$, therefore, the state is entangled only if $\lambda<\frac{3}{4}$.

\underline{Theorem}-Any 2-qubit, bipartite quantum state described by a rank 3 density operator is entangled only if the mixture has less than ${\frac{3}{4}}^{th}$ of the separable part. 

These bounds on entanglement are compatible with the results that have been previously discussed by Wunderlich, Plenio \color{blue}\cite{wunderich} \color{black}and Audenaert, Plenio \color{blue}\cite{audenaert} \color{black}in terms of local measurement of correlation observables in the canonical basis. It is straightforward to get the bound of Ineq. \color{blue}\ref{bound1} \color{black}in terms of observables, since $\lambda=\frac{3}{4}({\langle}s_zp_z{\rangle}+1)$, for $\rho^{max}_3$.

\section{Rank 4 states}
\label{sec:7}

\subsection{Bounds on concurrence}
In this most general case, we have absolutely no prior information on the state, and a complete tomography is inescapable. We may yet ask, as with rank 3 states, on the kinds of inferences that we may draw from partial information. To do so, we arrange the eigenexpansion
\begin{equation}
\rho_4=\lambda_1{\mathbb{I}_4}+\lambda_2\Pi_3+(1-\lambda_1-\lambda_2)\rho_2
\label{r4}
\end{equation}

Using the familiar resolution $\rho_2 =\mu\rho_{sep}+(1-\mu)|\psi\rangle\langle\psi|$, we see that  the only contribution to the entanglement of $\rho_4$ is from the pure state described by $|\psi\rangle$. The concurrence of $\rho_4$ is bounded by the relation
\begin{align}
\begin{split}
C(\rho_4)&\leq(1-{\lambda_1}-{\lambda_2})(1-\mu)C(\psi) \\
&\leq(1-\lambda_1-\lambda_2)C(\psi)
\end{split}
\label{ineq3}
\end{align}
Thus, $\rho_4$ is maximally entangled when, contribution of $\rho_{sep}$ in the mixture is zero and $|\psi\rangle$ is maximally entangled, i.e. $|\psi\rangle$ is a Bell state. In that case, we find that 
\begin{equation}
C(\rho^{max}_4)=\frac{\lambda_2}{3}ab+\frac{1-\lambda_1-\lambda_2}{2}-\frac{\lambda_1}{4}-\frac{\lambda_2}{3}
\label{c4}
\end{equation}
The above equations along with maximal entanglement condition $ab=\frac{1}{2}$, impose the constraint 

\begin{equation}
9\lambda_1+8\lambda_2<6
\label{ineq4}
\end{equation}
In addition to the normalization constraint $\lambda_1 + \lambda_2 \le 1$. Overall, the entangled and separable region in the parameter space is shown in Fig. \color{blue}\ref{fig:1}\color{black}.

We get back the rank-3 case for $\lambda_1=0$, which is consistent with the result.

\underline{Theorem}- Any 2-qubit, bipartite quantum state is entangled only if $\lambda_1$ and $\lambda_2$ as discussed above, which represent fractions of separable part in quantum state follow the relations,
\begin{align}
\begin{split}
9\lambda_1+8\lambda_2&<6 \\
\lambda_1+\lambda_2&\leq1
\end{split}
\label{r4f}
\end{align}

%
\begin{figure}
\label{fig:1}
\begin{center}
\resizebox{0.5\textwidth}{!}{%
  \includegraphics{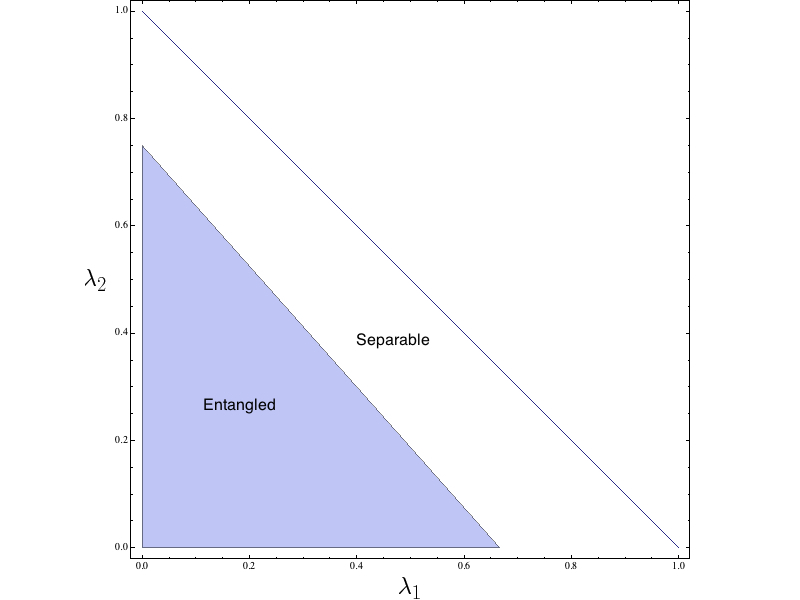}
}
\caption{Entanglement-Separability region for the rank-4 density matrix. The axes $\lambda_1$ and $\lambda_2$ represent fraction identity (completely mixed part) and rank-3 operator in the entanglement signal, respectively.}
\end{center}
\label{fig:1}       
\end{figure}

Finally, we conclude this section by exhibiting how correlation observables  determine the eigenvalues. Indeed for this maximally entangled state, a simple exercise shows that 
\begin{align*}
\begin{split}
\lambda_1&=1-{\langle}s_zp_z{\rangle}-2{\langle}s_xp_x{\rangle} \\
\lambda_2&=\frac{3}{2}({\langle}s_zp_z{\rangle}+{\langle}s_xp_x{\rangle})
\end{split}
\end{align*}

Above equations in combination with Ineq. \color{blue}\ref{r4f} \color{black}gives bounds in terms of correlation observables for this maximally entangled state. Entanglement of Werner state \color{blue}\cite{Barbieri} \color{black}lies in this class and satisfied by Ineq. \color{blue}\ref{r4f}\color{black}, which is more general since satisfied by even larger class of states.

\section{Conclusion}
The purpose is to extract maximal information about entanglement without exhausting complete set of observables while also considering the ease in measurement of certain observables over the other, for example local observables are easier to measure than correlations. In general the way a quantum state is prepared contains some information that can be used to restrict the parameter space on which entanglement depends. This can be vital in avoiding unnecessary measurements in entanglement experiments. We also highlight the invariant nature of entanglement in terms of other two qubit invariants, which depending on the system might be easier to determine experimentally.

We have shown for two qubit states when density matrices are rank 1(pure state) or rank 2, not only the entanglement can always be characterised without full quantum tomography, it can be expressed explicitly only using local observables. However, the rank 2 case comes with some exceptions, when the expectation value of one of the local observable becomes zero, we have to turn up to at least one nonlocal observable. We have pointed out one explicit example of such case and shown how to find its entanglement. For two qubit states described by rank 3 and rank 4 density matrices, because of more independent parameters, in general we require full quantum tomography. However, we have shown some specific cases when entanglement can be characterized with fewer observables. More importantly, we found loose bounds on entanglement in these cases. Since these bounds are functions of only one or two parameters, experimentally they have great potential in finding out if the quantum state is possibly entangled or not from very few observables.

\bigskip
\bigskip
\small
All authors contributed equally to this work.

\section*{Appendix}
\label{A}
The expectation value of correlation measurement on $\rho^{max}_3$ given by eqn.\ref{maxent1} is given by,
\begin{equation}
{\langle}s_zp_z{\rangle}=\frac{\lambda}{3}-(1-\frac{2\lambda}{3})+\frac{\lambda}{3}
\tag{A1}
\label{eqn:A1}
\end{equation}
This gives $\lambda$ in terms of correlation measurement. 

The expectation values of correlation measurements on $\rho^{max}_4$ are given by,
\begin{align}
\begin{split}
{\langle}s_xp_x{\rangle}&=\frac{2\lambda_2}{3}ab+1-\lambda_1-\lambda_2 \\
&=1-\lambda_1-\frac{2\lambda_2}{3} \\
{\langle}s_zp_z{\rangle}&=\frac{\lambda_1}{2}+\frac{2\lambda_2}{3}-\frac{\lambda_2}{3}-\frac{2-\lambda_1-2\lambda_2}{2} \\
&=\lambda_1+\frac{4\lambda_2}{3}-1
\end{split}
\tag{A2}
\label{eqn:A2}
\end{align}
Here, we have taken the maximally entangled case by substituting $ab=\frac{1}{2}$. The above equations can be solved to give $\lambda_1$ and $\lambda_2$ as functions of correlation measurements.

%
%
%

\end{document}